\documentclass{PoS}
\title{Towards a determination of $c_{SW}$ using Numerical Stochastic Perturbation Theory (NSPT)}

\ShortTitle{Towards a determination of $c_{SW}$ using NSPT}

\author{\speaker{C.~Torrero} and G.S.~Bali\\
        Institute for Theoretical Physics, University of Regensburg, 93040 Regensburg (Germany)\\
        E-mail: \email{christian.torrero@physik.uni-regensburg.de}
                \email{gunnar.bali@physik.uni-regensburg.de}}

\abstract{We outline a strategy to compute the second-loop contribution to the $c_{SW}$ coefficient of the Sheikoleslami-Wohlert-Wilson fermion action by means of NSPT. We also present preliminary results for higher-order integrators for the Langevin evolution within NSPT. At fixed numerical accuracy, these integrators considerably reduce the required computer-time.}

\FullConference{The XXVI International Symposium on Lattice Field Theory\\
		 July 14-19 2008\\
		 Williamsburg, Virginia, USA}

\begin{document}

\section{Motivation}

A problem one has often to face when handling lattice results is taking
the continuum limit.\\
\hspace*{0.7cm}A viable way of reducing the impact of lattice artifacts
(by \emph{removing} some of them) is given by the Symanzik improvement
programme~\cite{Sym} which, as is well-known, has allowed to remove
$\mathcal{O}(a)$ artifacts in unquenched simulations. The irrelevant term
to be added to the lattice action was determined by Sheikoleslami and
Wohlert~\cite{CSW} and contains the so-called $c_{SW}$ coefficient
which can be expanded perturbatively in even powers of the bare coupling
$g_{0}$.\\
While the zero- and one-loop coefficients of this expansion have already
been computed for different lattice actions~\cite{Woh,QCDSF} (see also
references therein), the two-loop contribution is still unknown:
within the Wilson formulation of Lattice QCD (LQCD), we address its
determination using NSPT, a tool that allows for perturbative
calculation in lattice-regularized quantum field theories.\\
\hspace*{0.7cm}In the second part, we discuss a technical issue
of NSPT that is related to the discretization of the Langevin equation
that governs the evolution of the system: our target is to obtain
high-order integrators, aiming at reducing the computer-time at fixed
accuracy.


\section{Wilson formulation of LQCD and improvement: some notation}

The Wilson action $S_W$ of LQCD can be decomposed into gauge ($S_G$) and
fermionic ($S_F$) parts with\footnote{In the equations of this and the
subsequent section it is understood that all dimensionful quantities
have been rescaled with powers of the lattice spacing $a$ to be
dimensionless (some of them carry an extra "hat" to emphasize this).},

\vspace*{-0.2cm}
\begin{equation} 
S_{G}=\beta\!\!\!\sum_{\scriptstyle n,~\!\!\mu,~\!\!\!\nu \atop \scriptstyle\mu~\!\!\!>\nu}\bigg(1-\frac{Tr}{2N_c}\big[U_{\mu\nu}(n)+{U}_{\mu\nu}^{\dagger}(n)\big]\bigg)~,
\end{equation}
\vspace*{-0.15cm}

\hspace*{-0.7cm}where $\beta = 2N_c/g_0^2$, $N_c$ is the number of colours,
$U_{\mu\nu}(n)$ is the lattice plaquette and,

\vspace*{-0.1cm}
\begin{equation}
S_{F}=\!\!\sum_{\scriptstyle n,~\!\!\alpha,~\!\!b \atop \scriptstyle m,~\!\!\beta,~\!\!c}\bar{\psi}_{\alpha b}(n)\mathcal{M}_{n\alpha b,\ \!m\beta c}[U]\psi_{\beta c}(m)\ ,
\end{equation}
\vspace*{-0.4cm}

\hspace*{-0.7cm}with
\vspace*{-0.5cm}

\begin{eqnarray}
\mathcal{M}[U]_{n\alpha b,~\!\!m\beta c}&=& -\frac{1}{2}\sum_{\mu}\bigg[\!(r-\gamma_{\mu})_{\alpha\beta}~\!U_{\mu}(n)^{\!}_{bc}~\!\delta_{n+~\!\!\hat{\mu},~\!\!m}+(r+\gamma_{\mu})_{\alpha\beta}~\!U_{\mu}^{\dagger}(m)^{\!}_{bc}~\!\delta_{n-~\!\!\hat{\mu},~\!\!m}\!\bigg] +\nonumber\\
&& +~\!(\widehat{M}_0+4r)~\! \delta_{nm}~\!\delta_{\alpha\beta}~\!\delta_{bc}\ .
\end{eqnarray}
\vspace*{-0.5cm}

\hspace*{-0.7cm}In eq.~(2.3),$~ r$ is the Wilson parameter (which we will set to $1$) while $\widehat{M}_0$ is the bare mass.\\
\hspace*{0.7cm}The Sheikoleslami-Wohlert irrelevant contribution ($S_{SW}$) to be added to the Wilson action is given by\footnote{In the following equations, spin and colour subscripts are suppressed to ease the notation: they can obviously be restored as in eq.~(2.3).},
\vspace*{-0.25cm}

\begin{equation}
S_{SW} = \frac{i}{4}~\!c_{SW}\!\!\!\sum_{n,\ \!\mu,\nu}\bar{\psi}(n)\sigma_{\mu\nu}\hat{F}_{\mu\nu}(n)\psi(n)\ ,
\end{equation}
\vspace*{-0.1cm}

\hspace*{-0.7cm}where $\sigma_{\mu\nu}=i/2[\gamma_{\mu},\gamma_{\nu}]$ while $\hat{F}_{\mu\nu}(n)$ reads,

\begin{equation}
\hat{F}_{\mu\nu}(n) = \frac{1}{8}\big(Q_{\mu\nu}(n)-Q_{\nu\mu}(n)\big) ,
\end{equation}
\vspace*{-0.25cm}

\hspace*{-0.7cm}with $Q_{\mu\nu}(n)$ being the clover term, $Q_{\mu\nu}(n) = U_{\mu,\nu}(n) + U_{-\nu,\ \!\mu}(n) + U_{\nu,-\mu}(n) + U_{-\mu,-\nu}(n)$.\\
\hspace*{0.7cm}As already anticipated, the coefficient $c_{SW}$ appearing in eq.~(2.4) can be decomposed as

\vspace*{-0.2cm}
\begin{equation}
c_{SW}  =  c_{SW}^{(0)} + c_{SW}^{(1)}g_0^2 + c_{SW}^{(2)}g_0^4 + \mathcal{O}(g_0^6)\ , 
\end{equation}
\vspace*{-0.4cm}
\
\hspace*{-0.7cm}where $c_{SW}^{(2)}$ is the target of our computation.


\section{How to measure $c_{SW}^{(2)}$}

A suitable observable to determine $c_{SW}^{(2)}$ is given by the
pion propagator\footnote{The subscript "U" means that the corresponding average has to be performed on gauge configurations only.},

\vspace*{-0.5cm}
\begin{eqnarray}
G_{bc}(n-m)&=&-\!\!\!\sum_{\scriptstyle \alpha,~\!\beta,~\!\delta,~\!\epsilon}\langle\bar{\psi}_{\alpha b}(n)(\gamma_5)_{\alpha\beta}\psi_{\beta b}(n)\bar{\psi}_{\delta c}(m)(\gamma_5)_{\delta\epsilon}\psi_{\epsilon c}(m)~\!\rangle = \nonumber\\
            &=&\sum_{\scriptstyle \alpha,~\!\beta,~\!\delta,~\!\epsilon}\langle(\gamma_5)_{\alpha\beta}\Big[\tilde{M}^{\ \!-1}\Big]_{n\beta b,\ \!m\delta c}(\gamma_5)_{\delta\epsilon} \Big[\tilde{M}^{\ \!-1}\Big]_{m\epsilon c,\ \!n\alpha b}~\!\rangle_U = \nonumber\\ 
            &=& \sum_{\scriptstyle \alpha,~\!\epsilon}\langle\Big[\tilde{M}^{\ \!-1}\Big]_{\!m\epsilon c,\ \!n\alpha b}^{*}\Big[\tilde{M}^{\ \!-1}\Big]_{m\epsilon c,\ \!n\alpha b}~\!\rangle_U
            = \sum_{\scriptstyle \alpha,~\! \epsilon}\langle\left|\Big[\tilde{M}^{-1}\Big]_{\!m\epsilon c,\ \!n\alpha b }\right|^2\rangle_U ,
\end{eqnarray}
\vspace*{-0.25cm}

\hspace*{-0.7cm}where $\tilde{M}$ is the fermionic operator obtained by adding
together eqs.~(2.2) and \nolinebreak (2.4): for details about its inversion,
see~\cite{DR0}.
After switching to momentum space and defining the dimensionless quantities $\hat{p}_{\mu}=p_{\mu}a$ and $\hat{p}^2=\sum_{\mu}\hat{p}_{\mu}^2$ (being the $p_\mu\!$'s the lattice momentum components), one can invert the propagator to obtain the $\hat{\Gamma}$-function which can be decomposed as,

\vspace*{-0.12cm}
\begin{equation}
\hat{\Gamma}(\hat{p},\hat{m}_{cr},g_0) = \hat{p}^2 + \widehat{M}_0^2 + \widehat{M}_W^2\!(\hat{p})- \hat{\Sigma}(\hat{p},\hat{m}_{cr},g_0)\ ,  
\end{equation}
\vspace*{-0.37cm}

\hspace*{-0.7cm}where $\widehat{M}_W(\hat{p})$ is the irrelevant Wilson mass,
$\widehat{M}_0$ the (perturbative) pion rest mass and $\hat{\Sigma}(\hat{p},\hat{m}_{cr},g_0)$ the self-energy with $\hat{m}_{cr}$ the critical mass defined as $\hat{m}_{cr} = \hat{\Sigma}(0,\hat{m}_{cr},g_0)$. Since we want to develop a mass-independent scheme, we will both set $\widehat{M}_0$ equal to zero and subtract the proper mass counterterms to keep fermions massless.\\
\hspace*{0.7cm}Given that we are eventually interested in a perturbative approach, we can expand $\hat{\Sigma}(\hat{p},\hat{m}_{cr},g_0)$ as a series in
$g_0^2$, i.e.\ $\hat{\Sigma}(\hat{p},\hat{m}_{cr},g_0) = \sum_{k}\hat{\Sigma}^{(k)}(\hat{p},\hat{m}_{cr})g_0^{2k}$, and decompose a generic coefficient \\ $\hat{\Sigma}^{(k)}(\hat{p},\hat{m}_{cr})$ by means of \emph{hypercubic invariants} as,

\vspace*{-0.1cm}
\begin{equation}
\hat{\Sigma}^{(k)}(\hat{p},\hat{m}_{cr}) = \alpha_1^{(k)}(\hat{m}_{cr}) + \alpha_2^{(k)}(\hat{m}_{cr})\sum_{\rho}\hat{p}_{\rho}^2 + \alpha_3^{(k)}(\hat{m}_{cr})\sum_{\rho}\hat{p}_{\rho}^4 + \ldots \ .
\end{equation}
\vspace*{-0.27cm}

A possible approach to determine $c_{SW}^{(2)}$ would consist of expanding
the pion and quark self energies\footnote{Formulae similar to
eqs.~(3.2)--(3.3) hold also for the quark propagator though the
Dirac structure is more involved.}
in a combined way: one could take $c_{SW}^{(0)}=1$ to obtain $\kappa_c$
to $\mathcal{O}(\beta^{-1}$), then take this
value to tune $c_{SW}^{(1)}$ to make the pion mass vanish; next,
one revisits the quark propagator to determine $\kappa_c$ to
$\mathcal{O}(\beta^{-2}$) and so on till $c_{SW}^{(2)}$ is
determined\footnote{It is clear that, requiring the pion to be massless,
also implies setting $\widehat{M}_0^2$ equal to zero.}. 

\hspace*{0.7cm}An alternative strategy could be the following:
recall that $S_{SW}$ was introduced to remove $\mathcal{O}(a)$
artifacts and the pion propagator contains the product of two quark
propagators. One should be able to establish a correspondance between
terms proportional to $a$ in the operator $\tilde{M}^{-1}$ and the
ones proportional to $a^2$ in the $\hat{\Gamma}$-function,
namely $\alpha_{3}^{(k)}$. If one now tunes $c_{SW}^{(0)}$ and
$c_{SW}^{(1)}$ to their known values and observes that, correspondingly,
$\alpha_{3}^{(1)}$ and $\alpha_{3}^{(2)}$ vanish, one can fix
$c_{SW}^{(2)}$ by requiring $\alpha_{3}^{(3)}$ to be zero. This
approach is maybe less rigorous but nonetheless should be worth studying.

\section{Numerical setup}

The method of our choice is NSPT. It is related to
\emph{Stochastic Quantization}~\cite{PW} which consists
of introducing an extra coordinate, a stochastic time $t$, 
together with an evolution equation of the Langevin type,

\vspace*{-0.20cm}
\begin{equation}
 \frac{\partial\phi(x,t)}{\partial t}=
  -\frac{\partial S[\phi]}{\partial\phi}+\eta(x,t)~,
\end{equation}
\vspace*{-0.28cm}
       
\hspace*{-0.7cm}where in this example $\phi(x,t)$ is a scalar field while $\eta(x,t)$ is a Gaussian noise.\\
Starting from this, the usual Feynman-Gibbs integration can be recovered by noise-averaging as

\begin{equation}
 Z^{-1}\!\!\int [D\phi]
 O[\phi\small(x\small)]e^{-S[\phi(x)]}=\lim_{t\rightarrow\infty}
 \frac{1}{t} \int_0^t \! {\rm d} t' \,
 \big\langle O[\phi_{\eta}\small(x,t'\small)]\big\rangle_{\eta}\;.
\end{equation}   
\vspace*{-0.25cm}

For $SU(3)$ lattice variables the Langevin equation has to
be modified into~\cite{DR0},

\vspace*{-0.10cm}
\begin{equation}
 \partial_{t}U_{\mu}(n,t) = -iT^A\Bigl( \nabla_{\!n,~\!\mu,~\!A} S[U]+\eta_{\mu}^A(n,t) \Bigr)
 U_{\mu}(n,t) \;,
\end{equation}
\vspace*{-0.30cm}

\hspace*{-0.7cm}in order to obtain an evolution of the variables within the group: here $T^A=\lambda^A/2$ are Gell-Mann matrices
while $\eta_{\mu}^A(n,t)$ are again Gaussian noise components.

The missing ingredient, i.e. Perturbation Theory, is introduced by expanding the $U$'s as\footnote{The expansion in the computer code is thought on $\beta^{-1/2}$ rather than $g_0$: converting the corresponding coefficients is obviously straightforward.},

\vspace*{-0.15cm}
\begin{equation}
 U_{\mu}(x,t)\longrightarrow\sum_{k}\beta^{-\frac{k}{2}}U_{\mu}^{(k)}(x,t) 
 \;,
\end{equation} 
\vspace*{-0.32cm}

\hspace*{-0.7cm}When plugging this into the Langevin equation,
this results in a system of coupled differential equations that
can be solved numerically via a discretization of the stochastic
time $t=N\tau$, where $\tau$ is a time step.\\
\hspace*{0.7cm}In practice, the system is evolved
for different values of $\tau$, then we average over each thermalized
signal to realize the limit $t\rightarrow \infty$ of eq.~(4.2).
Finally we extrapolate in $\tau$ to the $ \tau = 0 $ limit of the
desired observable: this extrapolation is needed since
\emph{the correct Boltzmann equilibrium distribution is
recovered only for continuous $t$}.


\section{High-order integrators for NSPT}

The $c_{SW}$-simulations are an ongoing project such that the
numerical results presented here refer to improvements of the NSPT
algorithm.\\
\hspace*{0.7cm}As mentioned above, one has to perform simulations
with different values of $\tau$ to extrapolate towards the limit
$\tau\rightarrow 0$ and this increases the required computer-time.
Since the smaller $\tau$ is, the more iterations
$N$ are needed, a possible way to save computer-time might
consist of employing larger values of $\tau$; however, this would 
compromise the accuracy of the subsequent $\tau$-extrapolation.
The solution to this drawback is well-known and is represented
by \emph{high-order integrators} for the Langevin equation:
these indeed increase the power of the leading $\tau$-dependence
thus allowing to safely recover the $\tau\rightarrow 0$ limit
even at large values of the time step.\\
\hspace*{0.7cm}The easiest way of determining high-order integrators
is probably by generalizing the usual Runge-Kutta schemes for the
scalar case: there, given a scalar variable $y(\tau)$, its derivative
$y'=f(\tau,y)$ and an initial value $y(\tau_0)\equiv y_0$,
the $m$-th order integrator reads:

\vspace*{-0.28cm}
\begin{equation}
y_{n+1} = y_n + \tau\sum_{l=1}^mb_lk_l \ \ \ \ \ \ \ \ \bigg(k_l \!=\! f\Big(\!\tau_n +c_l\tau, y_n\! +\! \tau\!\sum_{r=1}^{l-1}a_{l,r} k_r\!\Big)\ \ ; \ \ k_1 = f(\tau_n,y_n) \bigg)\ ,
\end{equation}
\vspace*{-0.13cm}

The generalization to non-Abelian variables appears straightforward:

\vspace*{-0.34cm}
\begin{eqnarray}
y_{n+1}\!\! = y_n + \tau\sum_{l=1}^mb_lk_l\!\! &\longrightarrow& \!\!  U_{\mu}(x,\tau_{n+1})\!=\!\exp\Big[\!\!-i\tau\!\!\sum_{l=1}^{m}\!b_l\!\Big(\eta_{\mu}(x,\tau_n)\!+\tilde{k}_l\Big)\!\Big]U_{\mu}(x,\tau_{n})\ \!, \\
\!\!\!\!\!\!\!\!\!\!\!\!\!\!\!\!\!\! k_l \!=\! f\Big(\!\tau_n\! +\!c_l\tau~\!,~\!y_n\! +\! \tau\!\sum_{r=1}^{l-1}\!a_{~\!\!l,~\!\!r} k_r\!\Big)\!\! &\longrightarrow& \!\!
\tilde{k}_l =\! \sum_{A}T^A\nabla_{x,~\!\!\mu,~\!\!A}S[~\widetilde{U}^{(l)}] \ \!, 
\end{eqnarray}
\vspace*{-0.16cm}

\hspace*{-0.7cm}where $S[\!\!~\widetilde{U}^{(l)}\ \!\!]$ is the expression of the action where all variables have been replaced as

\vspace*{-0.15cm}
\begin{equation}
U_{\mu}(x,\tau_n) \longrightarrow \exp\Big[\!\!-i\tau\!\sum_{r=1}^{l-1}a_{~\!\!l,~\!\!r}\Big(\eta_{\mu}(x,\tau_n)+\tilde{k}_r\Big)\Big]U_{\mu}(x,\tau_{n})\: ,
\end{equation}
\vspace*{-0.12cm}


\hspace*{-0.7cm}where it is understood that $\tilde{k}_1=\sum_{A}T^A\nabla_{x,~\!\!\mu,~\!\!A}S[U(\tau_n)]$. \\
As is manifest, the number of operations per update increases
with the order of the integrator: one will eventually be able to
employ larger time steps in the simulations, thus reducing the number
of iterations, but at the price of more costly iterations.
Our study seems to indicate that overall savings of up to a factor of
two can still be achieved.\\
\hspace*{0.7cm}Now everything boils down at getting the coefficients $a_{~\!\!l,~\!\!r}~\!\!, b_l, c_l$ in eq.~(5.1) but they can easily be found in the literature.\\ 
\hspace*{0.7cm}At present, the highest integrator available for the NSPT Langevin equation is second-order~\cite{Bat} and reads,

\vspace*{-0.6cm}

\vspace*{0.2cm}
    \begin{eqnarray}
     \!\!\!\!\!\!\!\!U_{\mu}(x,\tau_{n+1}) &=& exp\bigg[-i\bigg(1+\frac{\tau C_A}{6\beta}\bigg)\bigg(\frac{1}{2}\tau\tilde{k}_1 + \frac{1}{2}\tau\tilde{k}_2\bigg) - i\sqrt{\tau}\eta_{\mu}(x,\tau_n)\bigg]U_{\mu}(x,\tau_n) \ , \\
                                           & &  \nonumber\\
               \!\!\!\!\!\!\!\!\tilde{k}_1 &=&  \sum_A\!T^A\nabla_{x,~\!\!\mu,~\!\!A}S[U(\tau_n)]\ ,\\
                                           & &  \nonumber\\
               \!\!\!\!\!\!\!\!\tilde{k}_2 &=&  \sum_A\!T^A\nabla_{x,~\!\!\mu,~\!\!A}S[~\widetilde{\!U}^{(2)}]\ ,\\
                                           & &  \nonumber\\
     \!\!\!\!\!\!\!\!\widetilde{\!U}_{\mu}^{(2)}(x,.) &=&
exp\Big[-i\tau\tilde{k}_1 - i\sqrt{\tau} \eta_{\mu}(x,\tau_n)\Big]U_{\mu}(x,\tau_n)\ ,          
    \end{eqnarray}   

\hspace*{-0.7cm}where $C_A$ is the Casimir invariant of the Lie
group's adjoint representation: note that the noise employed
in eqs.~(5.5) and (5.8) is the same. The second term in the
first square brackets in eq.~(5.5) --- not appearing in Runge-Kutta
literature --- comes from the non-commutativity of group derivatives
introduced in eq.~(4.3). Apart from a rescaling\footnote{The
$\beta$ prefactor within eq.~(2.1) needs to be compensated for
and, consequently, the rescaling $\tau\mapsto\tau/\beta$
is required in \emph{every} integration scheme,
including Euler's most trivial one.} of $\tau$ with $\beta$~\cite{DR0},
the integrator can be determined by calculating the
equilibrium distribution of the corresponding discretized
\emph{Fokker-Planck equation}.\\
\hspace*{0.7cm}At present, analytical calculations are undertaken in order to compute the corresponding non-Abelian shifts for both the third- and the fourth-order integrator.     


\section{Preliminary results}

As can be seen from eq.~(5.5), the above-mentioned non-Abelian shift only affects loops higher than the first and thus we are already in the position of comparing one-loop results obtained with different integrators: the variable we choose is the plaquette at a lattice extent $L=4$.\\ Figure 1 shows how the slope in $\tau$ changes with the integrator as expected while, in Table 1, we collect our numerical results: as desired, the accuracy remains rather good even when employing larger $\tau$ values in the simulations. 

\vspace*{0.2cm}
\begin{table}[h]
\begin{center}
\begin{tabular}{|c|c|c|}
\hline
Order of integrator &   Employed time steps  &   1-loop plaquette  \\ 
\hline
        1           &  10,  15,  20  &  1.9930(7) \\
\hline
        2           &  50,  60,  70  &  1.9922(6) \\
\hline        
        3           &  90, 100, 110  &  1.9918(10) \\
\hline        
        4           &  110, 122, 130 &  1.9914(10) \\
\hline        
\end{tabular}
\vspace*{0.1cm}
\caption{Comparison between 1-loop results for different
integrators at L=4. the diagrammatic $L=4$ value reads
$1.9921875\ldots$.}
\label{Table 1}
\end{center}
\end{table}

\vspace*{-0.4cm}
The second-order integrator
should already be working to any perturbative order so that,
as a second test, we can check whether higher loops are
under control too when increasing the time steps in the simulations:
this is done in Table 2 where benchmark plaquette results at $L=4$
are provided by the first-order integrator that has been in
use since long.

\vspace*{0.2cm}
\begin{table}[h]
\begin{center}
\begin{tabular}{|c|c|c|c|c|c|}
\hline
Order of integrator &   1st loop  &   2nd loop   &   3rd loop  &   4th loop  \\ 
\hline
 1                  &  1.9930(7)  &  1.2027(18)  &  2.8781(67) &  8.994(30) \\
\hline
 2                  &  1.9922(6)  &  1.2002(17)  &  2.8778(62) &  8.990(28) \\
\hline 
\end{tabular}
\vspace*{0.1cm}
\caption{$\tau\rightarrow 0$ results from 1st and
2nd order integrators at $L=4$: the 1- and 2-loop diagrammatic
values
read $1.9921875\ldots$ and $1.2037037\ldots$.
The time steps employed can be found in Table 1.}
\label{Table 2}
\end{center}
\end{table}

\vspace*{-0.4cm}
In Figure~2 we compare the $\tau\rightarrow 0$ results
from the second order integrator to the corresponding
diagrammatic results
for different $L$. At $L=2$ there
is some disagreement which might be due to different ways of
treating zero modes. The ratios of diagrammatic
finite over infinite volume results read
0.907 (0.907) for $L=2$ and 0.994 (0.986) for $L=4$ at
1-loop (2-loop) level, respectively: finite volume effects are
much bigger than this disagreement between diagrammatic
results (neglecting zero modes) and NSPT (subtracting zero modes).
The infinite volume limit remains unaffected.


\section{Conclusions}

\vspace*{-0.1cm}
The computation of $c_{SW}^{(2)}$ is at an early stage: at present,
we are trying to single out the most reliable approach.\\
\hspace*{0.7cm}As for higher-order integrators for the Langevin equation,
first results seem to confirm good gains in computer-time, without
any loss in numerical accuracy.\\  


\vspace*{-0.58cm}

\section*{Acknowledgements}

\vspace*{-0.1cm}
We warmly thank the \emph{ECT*, Trento} for providing computer-time
on the \emph{BEN} cluster and the \emph{LRZ, Munich}.

\vspace*{0.93cm}


\vspace*{-7.5cm}
\begin{figure}[t]
 \vspace*{-0.7cm}
 \hfill
 \begin{minipage}[t]{.47\textwidth}
 \begin{center} 
   \hspace*{-0.5cm}%
   \includegraphics[width=6.7cm,height=6.2cm]{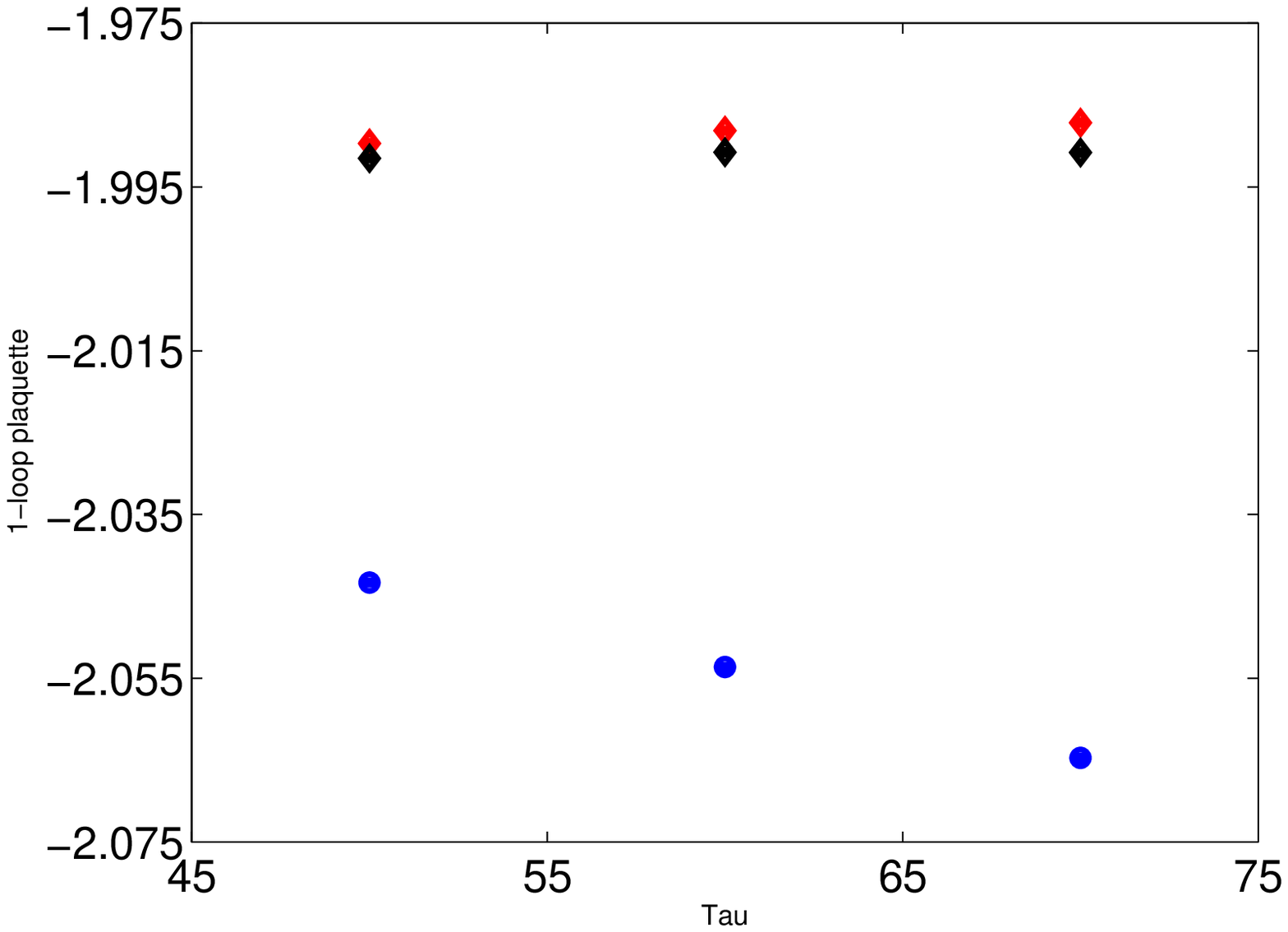}
   \vspace*{-0.45cm}
   \caption{1-loop plaquette vs. $\!\!\!\!\tau$ at L=4: data come from first-, second- and third-order integrator (blue, red and black points respectively).}
   \label{Fig.1}
  \end{center}
 \end{minipage}
 \hfill
 \begin{minipage}[t]{.47\textwidth}
  \begin{center} 
   \includegraphics[width=6.7cm,height=6.2cm]{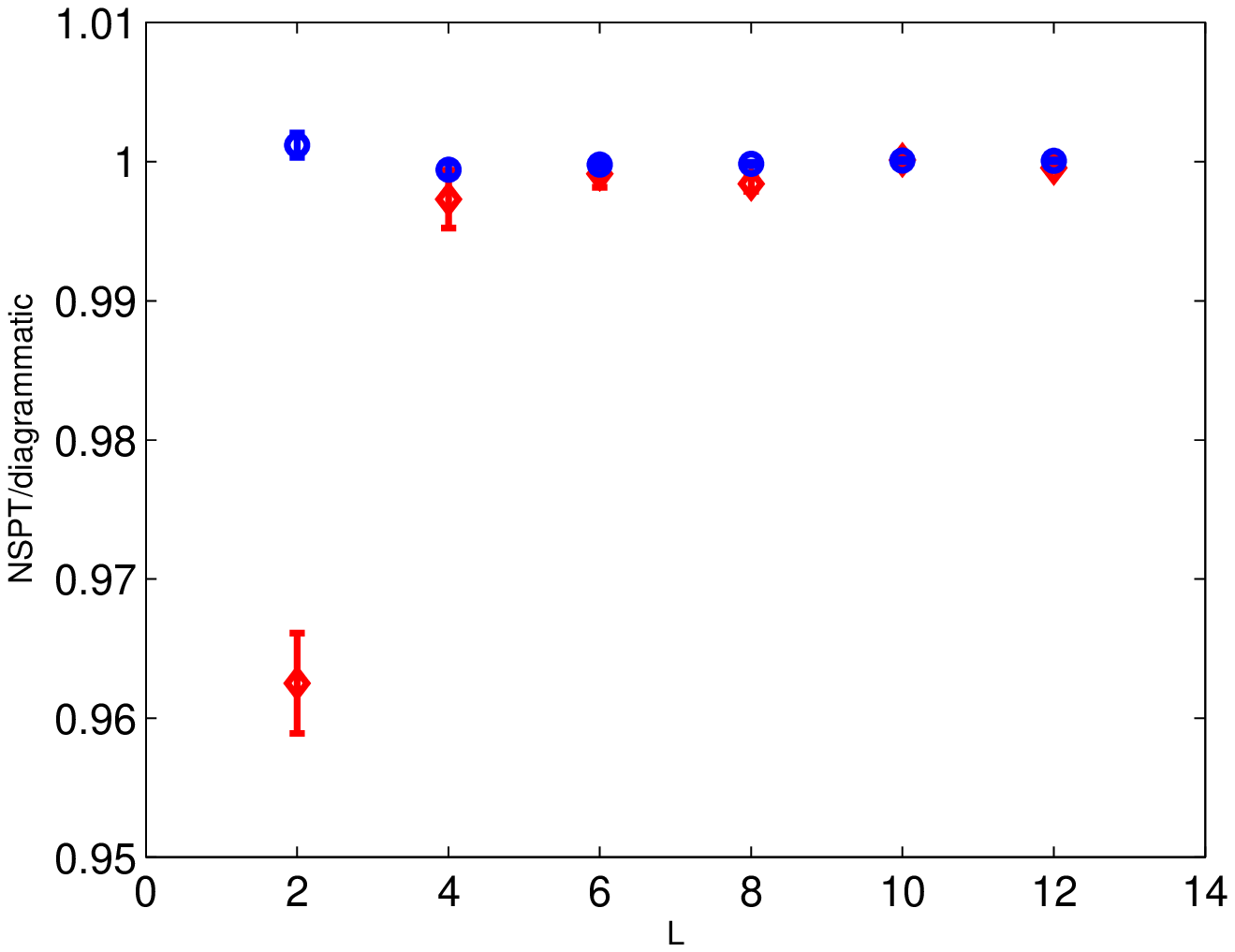}
   \vspace*{-0.45cm}
   \caption[a]{Ratio NSPT results/diagrammatic values vs. $\!L$ for the 1- and 2-loop plaquette (blue dots and red diamonds respectively).}
   \label{Fig.2}
  \end{center}
 \end{minipage}
 \hfill
\end{figure}


\end{document}